\documentclass[superscriptaddress,twocolumn,showpacs,prx,nofootinbib]{revtex4}

\newcommand{\beq}{\begin{equation}}
\newcommand{\eeq}{\end{equation}}
\newcommand{\bea}{\begin{eqnarray}}
\newcommand{\eea}{\end{eqnarray}}

\newcommand{\rme}{{\text e}}
\newcommand{\rmd}{{\text d}}

\newcommand{\ini}{\textrm{init}}
\newcommand{\fin}{\textrm{final}}
\newcommand{\upn}{\hat{n}}
\newcommand{\downn}{\hat{n}_\perp}
\def\bsigma{{\mbox{\boldmath $\sigma$}}}

\usepackage{graphicx,amsmath}
\usepackage{epstopdf}

\begin{document}

\title{Quantum Gates with Controlled Adiabatic Evolutions}
\author{Itay Hen}
\email{itayhen@isi.edu}
\affiliation{Information Sciences Institute, University of Southern California, Marina del Rey, California 90292, USA}

\date{\today}

\begin{abstract}
We introduce a class of quantum adiabatic evolutions that we claim may be interpreted as the equivalents of the unitary gates 
of the quantum gate model. We argue that these gates form a universal set and may therefore be used as building blocks 
in the construction of arbitrary `adiabatic circuits', analogously to the manner in which gates are used in the circuit model. 
One implication of the above construction is that arbitrary classical boolean circuits as well as gate model circuits 
may be directly translated to adiabatic algorithms with no additional resources or complexities. We show that while these adiabatic algorithms
fail to exhibit certain aspects of the inherent fault tolerance of traditional quantum adiabatic algorithms, 
they may have certain other experimental advantages acting as quantum gates. 
\end{abstract}

\maketitle

\section{\label{intro}Introduction}

Theoretical research on quantum computing is motivated by the exciting
possibility that quantum computers are inherently more efficient than
classical computers due to the advantages that the laws of quantum mechanics
provide, such as parallelism, tunneling and entanglement. The implications of having at our disposal reliable quantum computing devices with which diverse problems ranging from code breaking~\cite{shor:94} to database searching~\cite{grover:97}, are solved much faster 
than with the classical computers or supercomputers of today, are of course tremendous. 

The actual implementation of quantum computing devices is however hindered by many challenging difficulties, the most 
prominent  of which being the control or removal of quantum decoherence~\cite{schlosshauer:04}. 
Recent promising experimental research findings~\cite{mcgeoch:13,berkley:13,johnson:11} 
in the field of Adiabatic Quantum Computing (AQC) suggest that a leading candidate to be the first device to solve 
practical classically-hard problems using quantum principles is the so called `quantum annealer', 
which implements the simple yet potentially-powerful quantum-adiabatic algorithmic approach
proposed by Farhi {\it et al.}~\cite{farhi:01} about a decade ago.

The aforementioned experimental studies, as well as other theoretical work such as the theorem of polynomial
equivalence between AQC and the predominant gate model (GM) paradigm of quantum computing~\cite{aharonov:07,mizel:07}, provide ample motivation
for determining the computational capabilities of AQC and its precise relations with other 
quantum computing paradigms, specifically GM. Demonstrating that algorithms such as Shor's integer factorization~\cite{shor:94} are implementable 
as efficiently on a quantum adiabatic computer would undoubtedly have many practical as well as theoretical consequences
that would resonate well beyond Quantum Computing. 

Recent studies~\cite{roland:02,hen:14c,hen:14,hen:14b} examining the performance of certain AQC algorithms,
such as Unstructured Database Search~\cite{roland:02}, Quantum Counting~\cite{hen:14c} and Simon's problem~\cite{hen:14b},
against their GM counterparts, suggest that the equivalence between AQC and GM is stronger than the one implied by the principles
of polynomial equivalence prescribed in the seminal study of Aharonov {\it et al.}~\cite{aharonov:07}.
However, to date, no such strict equivalence or explicit construction to show that this is indeed the case, has been demonstrated. 

Here, we point out another connection between GM and AQC. We do this by considering the construction
of a class of quantum adiabatic algorithms, or subroutines, that we claim may be treated as the equivalents of the unitary gates 
of the quantum gate model. These `adiabatic gates' form a universal set and may therefore be used in the construction of 
general `adiabatic circuits', analogously to the manner in which gates are used in the circuit model. 
One implication of these constructions is that classical boolean circuits as well as gate model circuits 
may be directly translated to adiabatic algorithms with no additional resources or complexities, albeit without the
beneficial inherent robustness against dephasing that characterizes traditional AQC algorithms.

In our construction of adiabatic gates, we shall be using quantum adiabatic evolution
somewhat unconventionally. In our approach, 
we shall consider the adiabatic evolution of several systems in parallel. This approach will enable us to construct elaborate 
evolutions but at a cost, that we discuss later. 
The main principles of AQC as well as the new approach are presented next. 

\section{Controlled adiabatic evolution}

In AQC, one normally (albeit not exclusively) seeks the minimum value and corresponding input configuration of a given cost function, that is encoded as the final (or `problem') Hamiltonian, $\hat{H}^{(\textrm{f})}$,
such that the ground state of the final Hamiltonian and its energy are the solution to the original problem~\cite{farhi:01}. 
To find the solution, the system is prepared in the ground state of another `beginning' (or `driver') Hamiltonian
$\hat{H}^{(\textrm{b})}$ that must not commute with $\hat{H}^{(\textrm{f})}$ and has a ground state that is fairly easy to prepare. 
The Hamiltonian of the system is then slowly interpolated between $\hat{H}^{(\textrm{b})}$ and
$\hat{H}^{(\textrm{f})}$, normally via \hbox{$\hat{H}(t)=f_1(t) \hat{H}^{(\textrm{b})} +f_2(t) \hat{H}^{(\textrm{f})}$}
where $f_1(t)$ [$f_2(t)$] is a smoothly-varying function of time that is positive (zero) at $t=0$ and zero (positive) at $t=\mathcal{T}$. Here, $\mathcal{T}$ stands for the runtime of the algorithm. 
If this process is done slowly enough, the system will stay close to the ground state of the
instantaneous Hamiltonian throughout the evolution~\cite{kato:51,messiah:62}, so that one finally
obtains a state close to the ground state of $\hat{H}^{(\textrm{f})}$.  At this point,
measuring the state will yield the solution of the original problem with high
probability.  

It is clear from the above description, that the analog, continuous, nature of AQC is inherently very different from the discrete nature of GM
algorithms that are normally constructed by carrying out local unitary operations that act sequentially to advance the state of the system. 
For this reason it has been hard so far to draw meaningful analogies between AQC and GM. 
In what follows, we shall use a slightly unconventional `protocol' for adiabatic evolution, one which 
somewhat generalizes the above adiabatic procedure, and which, as we shall show, will allow us to perform more complicated calculations
than those allowed by the usual scheme.

Consider the following adiabatic-evolution Hamiltonian, defined over a bipartite system:
\bea \label{eq:Hfund}
\hat{H} &=& f_1(t) \cdot 1 \otimes \hat{H}^{(\textrm{b})} + f_2(t) \sum_j \hat{P}_j \otimes \hat{H}^{(\textrm{f})}_j \nonumber\\
&=& \sum_j \hat{P}_j \otimes \left[ f_1(t) \hat{H}^{(\textrm{b})} + f_2(t) \hat{H}^{(\textrm{f})}_j \right] \,,
\eea
where the operators $\{\hat{P}_j\}$ form a complete set of orthogonal projections on the first subsystem (i.e., \hbox{$\hat{P}_i \hat{P}_j=\hat{P}_i \delta_{ij}$} and $\sum_j \hat{P}_j=1$). The above Hamiltonian 
may be interpreted as one that executes a `controlled' adiabatic evolution of the second (target) subsystem, interpolating
between the beginning Hamiltonian $\hat{H}^{(\textrm{b})}$ and one of possibly several final Hamiltonians $\hat{H}^{(\textrm{f})}_j$,
the latter being determined by the state of the first (control) subsystem, via the projection operators.

For the initial state of the entire system $|\psi_\ini\rangle$ to be in the ground state of the total beginning Hamiltonian \hbox{$1 \otimes \hat{H}^{(\textrm{b})}$}, it suffices that $|\psi_\ini\rangle$ be in a product state $|\psi_\ini\rangle = |\psi\rangle \otimes |\textrm{g.s.}^{(\textrm{b})}\rangle$ where the state of the first subsystem $|\psi\rangle$ could be chosen arbitrarily, and the state of the second subsystem $|\textrm{g.s.}^{(\textrm{b})}\rangle$ is the 
(non-degenerate) ground state of $\hat{H}^{(\textrm{b})}$.

The choice as to which of the $\hat{H}^{(\textrm{f})}_j$ will serve as the final Hamiltonian, is determined by the state of the first subsystem.
If the state of the first subsystem $|\psi\rangle$ lies in the subspace projected by $\hat{P}_k$ for some $k$ (i.e., \hbox{$\hat{P}_j |\psi\rangle = \delta_{j k} |\psi\rangle$} for all $j$), then the final Hamiltonian will be $\hat{H}^{(\textrm{f})}_k$.
Of course, the state of the first system will in general have non-vanishing overlap on all subspaces projected by $\{ \hat{P}_j\}$. 
From simple linearity considerations, it is easy to see that in this general case, each of these components will evolve according to their 
respective final Hamiltonians, which would in turn mean that the total Hamiltonian will drive a number of independent adiabatic processes
in parallel, each corresponding to its own subspace in the Hilbert space of the first subsystem. Generally, controlled adiabatic evolution may be described as the process:
\beq
|\psi_\ini\rangle=|\psi \rangle |\textrm{g.s.}^{(\textrm{b})}\rangle \to 
|\psi_{\textrm{final}}\rangle=\sum_j \hat{P}_j|\psi\rangle  |\textrm{g.s.}_j^{(\textrm{f})}\rangle \,,
\eeq
where  $|\textrm{g.s.}_j^{(\textrm{f})}\rangle$ is the ground state of  $\hat{H}^{(\textrm{f})}_j$.
Note that in the trivial case where $\{\hat{P}_j\}=\{1\}$, the entire process is reduced to the usual adiabatic scheme. 

Utilizing the above form of adiabatic evolution, we will next demonstrate how one can use the principles of AQC to construct a class of quantum adiabatic algorithms,
which could be viewed as the equivalents of the gates of the gate model. Using these gates, as a direct consequence, 
adiabatic `circuits' may be constructed. The class of adiabatic gates that we consider below is the general single-qubit
rotation and its slightly more complicated generalization of controlled rotation.

\section{Adiabatic single-qubit rotation gates}

Consider a single qubit  
in an arbitrary unknown state $|\psi\rangle$.
Let us now attach to it an auxiliary qubit, initialized to the computational $|0\rangle$ state (which we shall identify as 
pointing in the positive $z$-direction):
\beq \label{eq:psib}
|\psi_\ini\rangle= |\psi \rangle \otimes |0\rangle \,.
\eeq
This will be the initial state of an adiabatic algorithm whose evolution will be governed by the Hamiltonian:
 \beq \label{eq:H}
\hat{H}(t)=|\upn\rangle \langle \upn| \otimes \hat{H}_{0}(t) + 
|\downn\rangle \langle \downn| \otimes \hat{H}_{\phi}(t) \,,
\eeq
where $\hat{H}_0(t)$ and $\hat{H}_\phi(t)$ are adiabatic-evolution Hamiltonians, conditioned to act only within the respective subspaces projected by 
the orthogonal projection operators \hbox{$|\upn\rangle \langle \upn| =1/2 \left( 1+\hat{n}\cdot \bsigma \right)$} and 
\hbox{$|\downn\rangle \langle \downn|=1/2 \left( 1-\hat{n} \cdot \bsigma \right)$} [where 
\hbox{$\bsigma=(\sigma_x,\sigma_y,\sigma_z)$}] defined on the Hilbert space of the first qubit. Here, $|\hat{n}\rangle$ 
and $|\hat{n}_\perp\rangle$ form a basis that corresponds to a predetermined unit vector $\hat{n}$ on the Bloch sphere of the first qubit.
The above Hamiltonian should be interpreted as driving two independently-acting, parallel, adiabatic processes [defined by $\hat{H}_0(t)$ and 
$\hat{H}_\phi(t)$], each acting within their own respective subspaces. 

The adiabatic-evolution Hamiltonians are chosen to be 
\beq \label{eq:Hphi}
\hat{H}_\phi(t)=-\cos \theta(t)  \sigma_z -\sin\theta(t) \left( \cos \phi \sigma_x + \sin \phi \sigma_y \right) \,,
\eeq
and \hbox{$\hat{H}_0(t) \equiv \hat{H}_{\phi=0}(t)=-\cos \theta(t)  \sigma_z -\sin\theta(t) \sigma_x$}. 
Similarly to $\hat{n}$, the angle $\phi$ is also a free parameter of the Hamiltonian.  
The time-dependence of the Hamiltonians 
is given here by the angle $\theta(t)$ such that $\theta(t=0)=0$, and \hbox{$\theta(t=\mathcal{T})=\theta_f$}, where $\theta_f$
is the value of the polar angle $\theta$ at the end of the evolution. For simplicity, we
shall henceforth assume the dependence \hbox{$\theta(t)=\theta_f \, t/\mathcal{T}$}. 

Note that the total Hamiltonian, Eq.~(\ref{eq:H}), is two-local and is of the general 
form introduced in Eq.~(\ref{eq:Hfund}), with \hbox{$\hat{H}^{(\textrm{b})}=-\sigma_z$} 
and \hbox{$\{ \hat{P}_1,\hat{P_2} \}= \{|\upn\rangle\langle \upn|,|\downn\rangle\langle \downn|\}$}. 
The two final Hamiltonians are \hbox{$\hat{H}^{(\textrm{f})}_1=-\cos \theta_f  \sigma_z -\sin\theta_f \sigma_x$} and \hbox{$\hat{H}^{(\textrm{f})}_2=-\cos \theta_f  \sigma_z -\sin\theta_f \left( \cos \phi \sigma_x + \sin \phi \sigma_y \right)$}.

\begin{figure}
\begin{center}
\includegraphics[width=5cm]{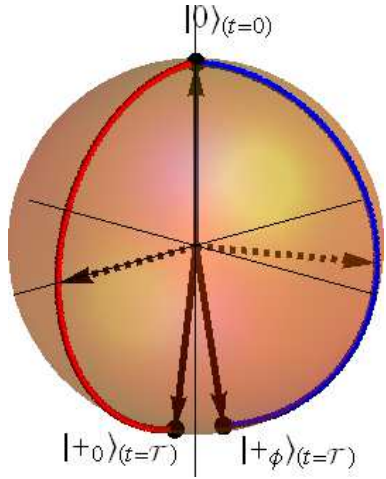}
\caption{Adiabatic evolution trajectories of the auxiliary qubit on the surface of the Bloch sphere during the adiabatic evolutions
$\hat{H}_0(t)$ and $\hat{H}_{\phi}(t)$. Starting at the $|0\rangle$ state, 
the state of the qubit `splits' into two trajectories. In the first it adiabatically evolves into the $|+_0\rangle$ state via the positive $x$-direction
(red dots) and in the 
second it evolves into $|+_\phi\rangle$ (blue dots) via the equator at angle $\phi$ from the positive $x$ axis. Both evolutions end at the same polar angle $\theta_f$ (which is close to $\pi$ in the figure).}
\label{fig:blochTraj}
\vspace{-0.7cm}
\end{center}
\end{figure}

Defining $|+_\phi\rangle \equiv \cos(\theta_f/2) |0\rangle + e^{i \phi} \sin(\theta_f/2) |1\rangle$, the Hamiltonian, Eq.~(\ref{eq:H}), 
will act differently and in parallel on the two complementary subspaces, evolving the auxiliary qubit, initially at $|0\rangle$, 
to $|+_{0}\rangle$ [the latter being the ground state of $\hat{H}_{0}(\mathcal{T})$] in the subspace 
projected by $|\upn\rangle \langle \upn| $ and to $|+_{\phi}\rangle$ [the ground state of $\hat{H}_{\phi}(\mathcal{T})$] in the subspace projected by 
$|\downn\rangle \langle \downn|$. 
The two evolutions are sketched in Fig.~\ref{fig:blochTraj}.

Writing the input qubit in the $\hat{n}$-basis as  
\hbox{$|\psi\rangle = \alpha |\hat{n}\rangle + \beta |\hat{n}_{\perp}\rangle$}, the final state of the system will be:
\bea \label{eq:psie}
|\psi_\fin\rangle &=& \alpha |\upn\rangle \otimes |+_{0} \rangle +
\beta |\downn\rangle \otimes  |+_{\phi} \rangle  \\\nonumber
&=&\cos(\theta_f/2) \left( \alpha |\upn\rangle + \beta |\downn\rangle  \right) \otimes |0\rangle \\\nonumber
&+&
\sin(\theta_f/2)  \left( \alpha |\upn\rangle +\rme^{i \phi} \beta |\downn\rangle  \right) \otimes |1\rangle\,.
\eea
It is crucially important to notice that the adiabatic evolution introduces no relative phase between the two end states
$|+_{0}\rangle$ and $|+_{\phi}\rangle$. This can be inferred directly from the
symmetries between the Hamiltonians $\hat{H}_0(t)$ and $\hat{H}_{\phi}(t)$. The parallel paths on the surface of the Bloch sphere 
traced by the auxiliary qubit are identical except they take place at different `longitudes' and so generate the same phases (see Fig.~\ref{fig:blochTraj})\footnote{In this simple case, the phases of the evolving states may be calculated analytically.  Each phase will have two contributions~\cite{polavieja:98}: The dynamic phase, given by 
$\theta(\mathcal{T})=\int_{0}^{\mathcal{T}} E(t') \rmd t'$ and the geometric (Berry) phase $\gamma(\mathcal{T}) =\arg\langle \psi(0) | \psi(T) \rangle+i \int_{0}^{\mathcal{T}} \langle \psi(t') | \dot{\psi}(t') \rangle \rmd t'$,
where $| \psi(t) \rangle$ is the evolving (time-dependent) eigenstate of the instantaneous Hamiltonian and $E(t)$ is its
instantaneous energy (in our units, $\hbar=1$). For the evolution described above, for both paths \hbox{$|0\rangle \to |+_{0}\rangle$} and \hbox{$|0\rangle \to |+_{\phi}\rangle$}, the instantaneous (ground state) energies $E(t)$ as well
as the overlap $\langle \psi(t') | \dot{\psi}(t') \rangle$ are identical throughout the evolution.}. 
The relative phase between the two final states therefore vanishes. 

Given the end state in Eq.~(\ref{eq:psie}), we see that in the limit of $\theta_f \to \pi$ the end state will be, with probability one, 
\hbox{$|\psi_{\textrm{final}}\rangle = |\psi_{\textrm{rot.}}\rangle \otimes |1\rangle$} where 
\beq
|\psi_{\textrm{rot.}}\rangle =  \alpha |\upn\rangle +\rme^{i \phi} \beta |\downn\rangle \,,
\eeq
i.e., the first qubit ends up rotated by an angle $\phi$ around the $\hat{n}$-axis.

We therefore now have an adiabatic machinery to rotate a qubit by an arbitrary angle $\phi$ around an arbitrarily chosen axis $\hat{n}$, i.e., we have constructed an adiabatic, general, single-qubit `gate'.

\section{Adiabatic controlled-rotation gates}

The above scheme may now be easily generalized to the case where the system initially contains two input qubits, one of which 
is regarded as a control qubit. 
Here, the starting state would be the two-qubit state 
\hbox{$|\psi\rangle = \alpha |0 ,\upn\rangle + \beta |0 ,\downn\rangle + \gamma |1 ,\upn\rangle + \delta |1,\downn\rangle$}, where the first qubit here is used as the control qubit. An adiabatic controlled rotation is obtained by attaching, as before,
an auxiliary qubit to the initial state:
\beq
|\psi_\ini\rangle= |\psi \rangle \otimes |0 \rangle \,,
\eeq
and constructing the slightly more complicated three-local Hamiltonian\footnote{
One could imagine replacing the three-local Hamiltonian with an equivalent perturbative gadget consisting of only two-local 
interactions but yielding the same effect, in order to make the model more attractive experimentally~\cite{davePrivate}. This could be done 
by employing the gadget proposed by Kempe, Kitaev, and Regev~\cite{kempe:06}, by which arbitrary three-body effective interactions can be obtained using Hamiltonians consisting only of two-body interactions.}:
\bea
\hat{H}(t) = |1,\downn\rangle \langle 1,\downn| \otimes \hat{H}_{\phi}(t)\\\nonumber
+\left( |0,\upn\rangle \langle 0,\upn| + |0,\downn\rangle \langle 0,\downn|+
|1,\upn\rangle \langle 1,\upn|\right) \otimes \hat{H}_{0}(t)
\,,
\eea
where $\hat{H}_0(t)$ and $\hat{H}_{\phi}(t)$ are as previously defined. 
In the limit of $\theta_f \to \pi$, the end state in this case will be, the controlled-rotated state \hbox{$|\psi_{\textrm{c. rot.}}\rangle \otimes |1\rangle$} where:
\bea
|\psi_{\textrm{c. rot.}}\rangle &=& \left( \alpha |0 ,\upn\rangle + \beta |0 ,\downn\rangle \right) 
\\\nonumber &+&
\left(  \gamma |1 ,\upn\rangle + \rme^{i \phi} \delta |1,\downn\rangle \right) \,.
\eea

The combination of the controlled gate above and the single-qubit general-rotation gate described earlier,
suggest that we now have in our possession a universal set of gates, with which general `adiabatic circuits' may be built, 
analogously to the manner in which algorithms are constructed in the gate model. Note that while here the controlled rotation gate 
has been chosen as our two-qubit gate,  any other nontrivial two-qubit gate, when combined with a universal one-qubit gate, would have yielded a universal set of gates. Moreover, it should be clear that the 
above scheme for the controlled rotation gate can be easily generalized to yield a general two-qubit gate.

It would be advantageous though to explicitly describe a couple of specific gates.
The NOT gate, for example, corresponds to a rotation by $\pi$ around the $x$-axis, i.e., it is the single-qubit rotation gate with 
the choice $\phi=\pi$ and $|\hat{n}\rangle = |+\rangle$:
\bea
\alpha|0\rangle + \beta |1\rangle &=& \frac{\alpha+\beta}{\sqrt{2}}|+\rangle +
\frac{\alpha-\beta}{\sqrt{2}}|-\rangle\\ \nonumber
 &\overrightarrow{\pi \, \textrm{rot.}}& \frac{\alpha+\beta}{\sqrt{2}}|+\rangle -
\frac{\alpha-\beta}{\sqrt{2}}|-\rangle = \beta|0\rangle + \alpha |1\rangle \,,
\eea
where $|\pm\rangle = \frac1{\sqrt{2}}(|0\rangle \pm  |1\rangle)$. The `controlled' version of this gate, namely CNOT,
may be just as easily constructed, using the controlled-rotation adiabatic scheme. 

Another example, is the Hadamard adiabatic gate which is simply a $\pi/2$ rotation around the $y$-axis:
\bea
\alpha|0\rangle + \beta |1\rangle &=& \frac{\alpha-i\beta}{\sqrt{2}}|+_y\rangle +
\frac{\alpha+i \beta}{\sqrt{2}}|-_y\rangle \\\nonumber
&\overrightarrow{\pi/2 \, \textrm{rot.}}&
\frac{\alpha-i\beta}{\sqrt{2}}|+_y\rangle +i
\frac{\alpha+i\beta}{\sqrt{2}}|-_y\rangle \nonumber\\
&=& \frac{\alpha+\beta}{\sqrt{2}}|0\rangle +\frac{\alpha-\beta}{\sqrt{2}} |1\rangle\,, \nonumber
\eea
where $|\pm_y\rangle = \frac1{\sqrt{2}}(|0\rangle \pm i |1\rangle)$ and we have omitted the immaterial global phase.

\section{Adiabatic Quantum Circuits}

It should now be clear that the class of general single-qubit and controlled rotation gates proposed above
are universal, and so general adiabatic circuits may be constructed using sequences of those. 

First, we note that while the above adiabatic `gates' were shown to act on isolated qubits, the linearity of Quantum Mechanics 
ensures that the above scheme holds even if the target qubits are part of a larger system of qubits in a more complicated state. 

A sequence of such gates in the above form, that are turned slowly on and off, may thus be used, one after the other to 
form quantum (as well as classical) circuits  similarly to the manner in which circuits 
are constructed in the usual gate model, this time only using concatenated purely-adiabatic evolutions.   
The final state of the adiabatic evolution of one gate would serve as the initial state of the next gate in the sequence. 
The standard initial state of the adiabatic circuit, in which all qubits are set to the computational $|0\rangle$ state, can also be easily prepared by applying an appropriate longitudinal magnetic field that is turned off once the first gate in the circuit is applied. Within the above scheme, one would in principle need one auxiliary qubit for each gate in the circuit although it should be clear that gates that act at different times slices may utilize the same auxiliary qubit as their ancillary resource.


The adiabatic gates proposed above are based on the concept of `controlled adiabatic evolution' introduced above, which describes several independently-evolving adiabatic processes. 
Each such process is a simple evolution of one qubit on the surface of a Bloch sphere. As can be easily inferred by looking at the adiabatic-evolution Hamiltonians, Eq.~(\ref{eq:Hphi}), the gap in these  adiabatic evolutions 
is constant throughout the evolution (and equals to $2$). Therefore, the required runtime for each adiabatic gate scales with neither the total number of qubits in the system nor with the number of gates in the circuit. The total runtime of a circuit of $S$ basic gates is therefore simply $O(S)$. 

An important remark is now in order. Our newfound ability to rotate a qubit using purely adiabatic evolutions comes at a cost. The independently evolving processes that yield the adiabatic gates have ground state manifolds that are doubly-degenerate. This is in contrast with traditional AQC setups in which the ground state is uniquely defined. The distinction between these two cases is important mainly because it is this uniqueness that normally provides AQC with the attractive property of being robust (to the extent that it is) against the devastating effects of decoherence, unlike other paradigms of quantum computation~\cite{childs:01,amin:09a}. The doubly-degenerate ground state manifolds of the adiabatic gates suggest that, while very versatile, they are likely to be more vulnerable to the effects of noise, similarly to the situation that arises in holonomic quantum computation~\cite{zanardi:99,carollo:05} and adiabatic gate teleportation~\cite{bacon:09,bacon:10}. It is important to note however that while the degeneracy of the ground state makes the method susceptible to errors of dephasing in the energy eigenbasis, it still enjoys a (constant) gap which protects it from the other forms of errors.  

In addition, even though the present method does not possess all the natural robustness of AQC, degenerate ground state quantum computation may certainly benefit from other types of fault tolerance schemes (see, e.g., Refs.~\cite{oreshkov:09,lidar:13}). Moreover, it is worth mentioning that the fact that adiabatic algorithms constructed via the method presented here consist of gates, advantageously allows for the utilization of gate-model error correction schemes and principles. The present method can thus be viewed as combining advantages of the gate model, specifically modularity, with some of the inherent robustness of AQC.   

\section{Summary and conclusions} 

We have shown how to use controlled adiabatic evolutions to construct general single-qubit and controlled two-qubit `adiabatic gates'
that can further be used as building blocks in the construction of general, arbitrary quantum circuits (as well as classical boolean circuits) 
in a straightforward manner. These evolutions possess the simple geometric representations of parallely-evolving paths on the surface of the Bloch sphere with vanishing geometric phases. Moreover, we have demonstrated that the construction of such adiabatic circuits comes at no additional complexity cost or resource overhead. For example, one could straightforwardly construct an adiabatic version of Shor's integer factorization algorithm~\cite{shor:94} using only two-local and three-local Hamiltonians for its adiabatic gates. The theoretical and practical implications of an implementable Shor's algorithm, on a many-qubit quantum annealer that will become available in the near future~\cite{mcgeoch:13,berkley:13,johnson:11}, may be tremendous,  both in the field of Quantum Computing and well beyond it. 

Adiabatic quantum computing, in its traditional form, has been shown 
to have several advantages over the gate model~\cite{childs:01,amin:09a} making it more fault-tolerant and robust 
against decoherence and dephasing. However the circuit adiabatic-evolution construction proposed here differs from traditional AQC
in two main features. First, the usual AQC is normally thought of as one continuous process 
interpolating between one beginning Hamiltonian and one final Hamiltonian, thereby 
eliminating the need for gates, that usually also carry around gate errors and therefore need 
error correction. Second, within the usual AQC scheme, the existence of a gap between the ground state and the rest of the spectrum throughout the adiabatic evolution serves to protect the system against decoherence and dephasing.

As discussed in the previous section, the method proposed above utilizes adiabatic gates as well as degenerate ground states, which seemingly implies
a lack of the natural AQC robustness. It is therefore important to note that while the existence of a degenerate ground subspace implies the lack of robustness against some types of errors (namely, dephasing in the energy eigenbasis), the constant gap separating this subspace from excited states, grants it an inherent fault tolerance against other types of errors. In addition, the usage of adiabatic evolutions as gates gives the scheme presented here
 the modularity of the gate model which further enables the construction of complicated algorithms as circuits as well as the ability to employ the standard methods of gate-model error correction. The present method can thus be considered as a hybrid between AQC and GM, combining some of the advantages of the two paradigms.

It would be interesting to know whether the adiabatic gates presented here are more amenable to
fault-tolerant types of error corrections when compared against unitary non-adiabatic evolution gates. Recent experimental evidence~\cite{martinis:14} demonstrating a controlled phase-shift gate relying on adiabatic interactions in superconducting Xmon transmon qubits with very high fidelities,  suggests that adiabatic gates may certainly be more powerful in practice than non-adiabatic ones. 
A very recent theoretical work~\cite{wiebe:12,kieferova:14} has illustrated that the ideas of controlled adiabatic evolution may become 
advantageous in adiabatic state preparation and other important scenarios. These and other promising ideas still remain be fully explored.

\begin{acknowledgments}
We thank Mohammad Amin, Dave Bacon, Amir Kalev, Daniel Lidar, Eleanor Rieffel, Federico Spedalieri and Peter Young for useful comments and discussions.
This project was supported by ARO grant number W911NF-12-1-0523. 
\end{acknowledgments}

\bibliography{refs,notes}



\end{document}